\documentclass[alpha-refs]{nbdt-article}

\usepackage{natbib}
\usepackage{siunitx}
\usepackage{setspace}
\usepackage{hyperref}
\usepackage[justification=justified,singlelinecheck=false]{caption}

\papertype{Original Article}
\paperfield{Neural Data Science/Analysis}
\jname{Neurons, Behavior, Data Analysis and Theory}
\jyear{2021}
\jpages{1-18}
\jvolume{5(3)}
\jwebsite{https://nbdt.scholasticahq.com/}
\paperdoi{\href{http://dx.doi.org/10.51628/001c.27680}{10.51628/001c.27680}}
\paperreceived{13th January 2021}
\paperrevised{15th June 2021}
\paperaccepted{5th July 2021}

\title{Statistical analysis of periodic data in neuroscience}

\author[1]{Daniel H. Baker}

\affil[1]{Department of Psychology and York Biomedical Research Institute, University of York, United Kingdom}

\corraddress{Department of Psychology, University of York, Heslington, York, YO10 5DD, United Kingdom}
\corremail{daniel.baker@york.ac.uk}

\fundinginfo{This research was not supported by external funding}

\runningauthor{Baker, D.H.}

\begin{document}
\maketitle

\begin{abstract}
\begin{spacing}{0.85}Many experimental paradigms in neuroscience involve driving the nervous system with periodic sensory stimuli. Neural signals recorded using a variety of techniques will then include phase-locked oscillations at the stimulation frequency. The analysis of such data often involves standard univariate statistics such as T-tests, conducted on the Fourier amplitude components (ignoring phase), either to test for the presence of a signal, or to compare signals across different conditions. However, the assumptions of these tests will sometimes be violated because amplitudes are not normally distributed, and furthermore weak signals might be missed if the phase information is discarded. An alternative approach is to conduct multivariate statistical tests using the real and imaginary Fourier components. Here the performance of two multivariate extensions of the T-test are compared: Hotelling's \(T^2\) and a variant called \(T^2_{circ}\). A novel test of the assumptions of \(T^2_{circ}\) is developed, based on the condition index of the data (the square root of the ratio of eigenvalues of a bounding ellipse), and a heuristic for excluding outliers using the Mahalanobis distance is proposed. The \(T^2_{circ}\) statistic is then extended to multi-level designs, resulting in a new statistical test termed \(ANOVA^2_{circ}\). This has identical assumptions to \(T^2_{circ}\), and is shown to be more sensitive than MANOVA when these assumptions are met. The use of these tests is demonstrated for two publicly available empirical data sets, and practical guidance is suggested for choosing which test to run. Implementations of these novel tools are provided as an \emph{R} package and a \emph{Matlab} toolbox, in the hope that their wider adoption will improve the sensitivity of statistical inferences involving periodic data.


\end{spacing}

\end{abstract}

\hypertarget{background}{%
\section{Background}\label{background}}

A widely used approach in many branches of neuroscience is to drive the nervous system using periodic stimuli. This entrains neural responses at the stimulation frequency, resulting in high signal-to-noise ratios relative to single stimulus presentations. These periodic responses, often called the \emph{steady-state} or \emph{frequency following} response, can be recorded using invasive methods from single neurons \citep{Enroth-Cugell1966} and local field potentials \citep{Morrone1987}, or with non-invasive electroencephalography (EEG) and magnetoencephalography (MEG) systems, both in humans \citep{Norcia2015} and in diverse animal species including insects \citep{Afsari2014}, birds \citep{Porciatti1990}, rodents \citep{Hwang2019} and primates \citep{Nakayama1982}. Steady-state methods are used to measure early sensory responses in vision \citep{Regan1966}, hearing \citep{Rees1986} and somatosensation \citep{Snyder1992}, and closely related paradigms have been developed to target specific stimulus features such as orientation \citep{Braddick1986}, and facial expression \citep{Gray2020} and identity \citep{Liu-Shuang2014}. In fMRI research, \emph{travelling wave} methods \citep{Engel1994, Sereno1995} are used to map the retinotopic responses of early visual cortex using stimuli that change periodically in spatial position. Finally, physiological reflexes such as the pupillary response to light can be entrained in a similar way \citep{Spitschan2014}.

\begin{figure}

{\centering \includegraphics[width=0.9\textwidth]{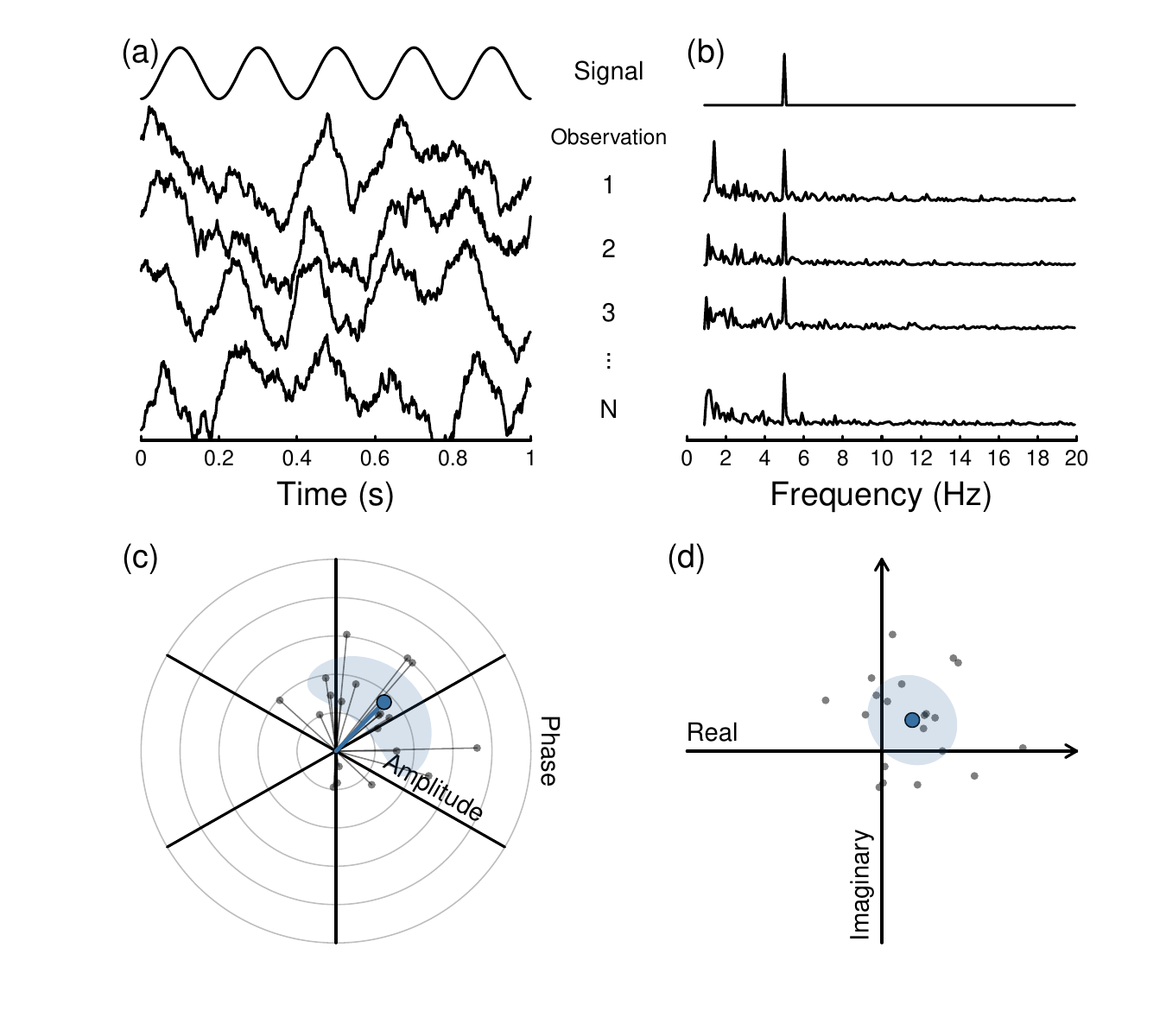} 

}
\caption{Illustration of the principles of Fourier analysis for time-varying signals. Panel (a) shows a sinusoidal signal waveform, and simulated neural responses for successive observations (repetitions in the same individual, or recorded from multiple individuals). Panel (b) shows the Fourier amplitude spectra of the waveforms in (a), with clear peaks at the signal frequency (5Hz) in each example. The Fourier spectrum also includes a phase term, which can be represented in polar coordinates (amplitude and phase); panel (c) shows this for example responses at the signal frequency (grey points) and their average (blue point), which is computed separately for amplitude and phase terms. Alternatively, the same information is contained in a Cartesian representation of the real and imaginary parts of the complex spectrum (panel (d)). Notice that the individual observations (grey points) are the same in panels (c,d), but the different representations have implications for how the average (blue points) and measures of spread (shaded regions) are calculated.}\label{fig:fourierexplain}

\end{figure}

A convenient way to analyse the data from periodic stimulation experiments is to take the Fourier transform of the measured signal. The amplitude of the response at the stimulation frequency (and its harmonics - integer multiples of the stimulation frequency) is a precise and well-isolated index of the brain's response (see Figure \ref{fig:fourierexplain}a,b). Fourier spectra comprise both amplitude and phase information that can be expressed in polar coordinates (Figure \ref{fig:fourierexplain}c), or equivalently as complex numbers with real and imaginary components (Figure \ref{fig:fourierexplain}d). In many studies the phase information is routinely discarded, and statistical comparisons are performed on the amplitude data (or equivalently power or signal-to-noise ratios) only \citep[e.g.][note this is a non-exhaustive selection of recent studies]{Liu-Shuang2014, Afsari2014, Hou2020, Itthipuripat2014, McFadden2014, Smith2017, Vanegas2015}. Typically researchers wish to test the experimental hypothesis that a signal is present against the null hypothesis that it is absent, or they wish to know if responses differ in amplitude and/or phase between two (or more) experimental conditions. However, there are several objections to using univariate statistics to answer these questions, as will be demonstrated below. An alternative is to use multivariate statistics, which take into account both the amplitude and phase information (represented as real and imaginary Fourier components in a Cartesian space). Multivariate methods have the advantage that they are more sensitive to weak signals, and therefore offer increased statistical power relative to univariate methods for typical applications (detecting the presence of a signal, or comparing multiple signals).

For pointwise and pairwise comparisons, Hotelling's \(T^2\) statistic \citep{Hotelling1931} is a multivariate extension of the T-test. For the one-sample case, the test statistic is defined as:
\begin{equation}
\label{eq:t2eq}
T^2 = N(\bar{x} - \mu)' C^{-1} (\bar{x} - \mu),
\end{equation}
where \emph{N} is the number of observations, \(\bar{x}\) is the multivariate sample mean, \(\mu\) is the point of comparison, \(C^{-1}\) is the inverse covariance matrix, and \('\) denotes vector transposition. Conceptually, the \(T^2\) statistic extends the univariate T-statistic by incorporating the covariance between the dependent variables. Two-sample and paired variants are also available, and the test can be applied with an arbitrary number of dependent variables (though here only the bivariate case will be considered).

More recently, \cite{Victor1991} proposed a simpler version of \(T^2\), called \(T^2_{circ}\). The \(T^2_{circ}\) statistic makes the strong assumption that the dependent variables (real and imaginary Fourier components) are uncorrelated and have equal variance. When these conditions are met, the test statistic for the one-sample case is defined as:
\begin{equation}
\label{eq:t2c}
T^2_{circ} = (N-1)\frac{|\bar{x}-\mu|^2}{\Sigma|x_j - \bar{x}|^2},
\end{equation}
where \(x_j\) denotes the \(j\)th observation of the dependent variables, and all other terms retain their previous meanings. Notice that no covariance term is present in equation \eqref{eq:t2c}, because of the independence assumption. This makes the statistic simpler to calculate, but causes problems when the assumption is violated (as will be demonstrated below). Conceptually, this statistic takes the vector difference between the bivariate sample mean (\(\bar{x}\)) and a comparison point (\(\mu\)), and scales by the mean length of the residual vector lines joining each data point to the sample mean. Two-sample and repeated measures versions of the \(T^2_{circ}\) statistic are also possible.

In the present paper, best practice guidelines are developed for performing statistical tests on multivariate Fourier components derived from periodic stimulation paradigms. It is first demonstrated why parametric univariate statistics are inappropriate for such data, primarily because amplitudes for weak signals are not normally distributed. Then conditions are investigated under which either the \(T^2\) or \(T^2_{circ}\) statistic should be used. The range of sample sizes and effect sizes where \(T^2_{circ}\) is more sensitive is identified. A novel method for testing the assumptions of the \(T^2_{circ}\) statistic is developed, based on calculating the condition index of a multivariate data set. Appropriate methods for identifying outliers using the Mahalanobis distance are discussed, and a heuristic proposed. Next the logic of \(T^2_{circ}\) is extended to situations with more than two levels of the independent variable, and the performance of this novel \(ANOVA^2_{circ}\) statistic is compared to MANOVA. Finally, the proposed techniques are demonstrated on two example data sets (from mice and humans), and best practice guidelines are recommended for analysis decisions.

All scripts used to generate this manuscript are available at: \url{https://github.com/bakerdh/FourierStats}. This includes a \emph{Matlab} toolbox and an \emph{R} package called \emph{FourierStats}, featuring functions to implement Hotelling's \citeyearpar{Hotelling1931} \(T^2\) statistic, Victor \& Mast's \citeyearpar{Victor1991} \(T^2_{circ}\) statistic, and the condition index and \(ANOVA^2_{circ}\) statistics proposed in this paper.

\hypertarget{fourier-amplitudes-violate-parametric-assumptions-of-univariate-statistics}{%
\section{Fourier amplitudes violate parametric assumptions of univariate statistics}\label{fourier-amplitudes-violate-parametric-assumptions-of-univariate-statistics}}

Many empirical studies use univariate T-tests or analysis of variance (ANOVA) to analyse periodic data. Specifically, the amplitude component of the Fourier spectrum at the stimulation frequency is used as the dependent variable, discarding the phase information. This is problematic, because the amplitude is an absolute quantity, and can never fall below zero. Distributions of amplitudes for weak signals are therefore positively skewed, and will generally violate the assumption of normality.

\begin{figure}

{\centering \includegraphics[width=0.95\textwidth]{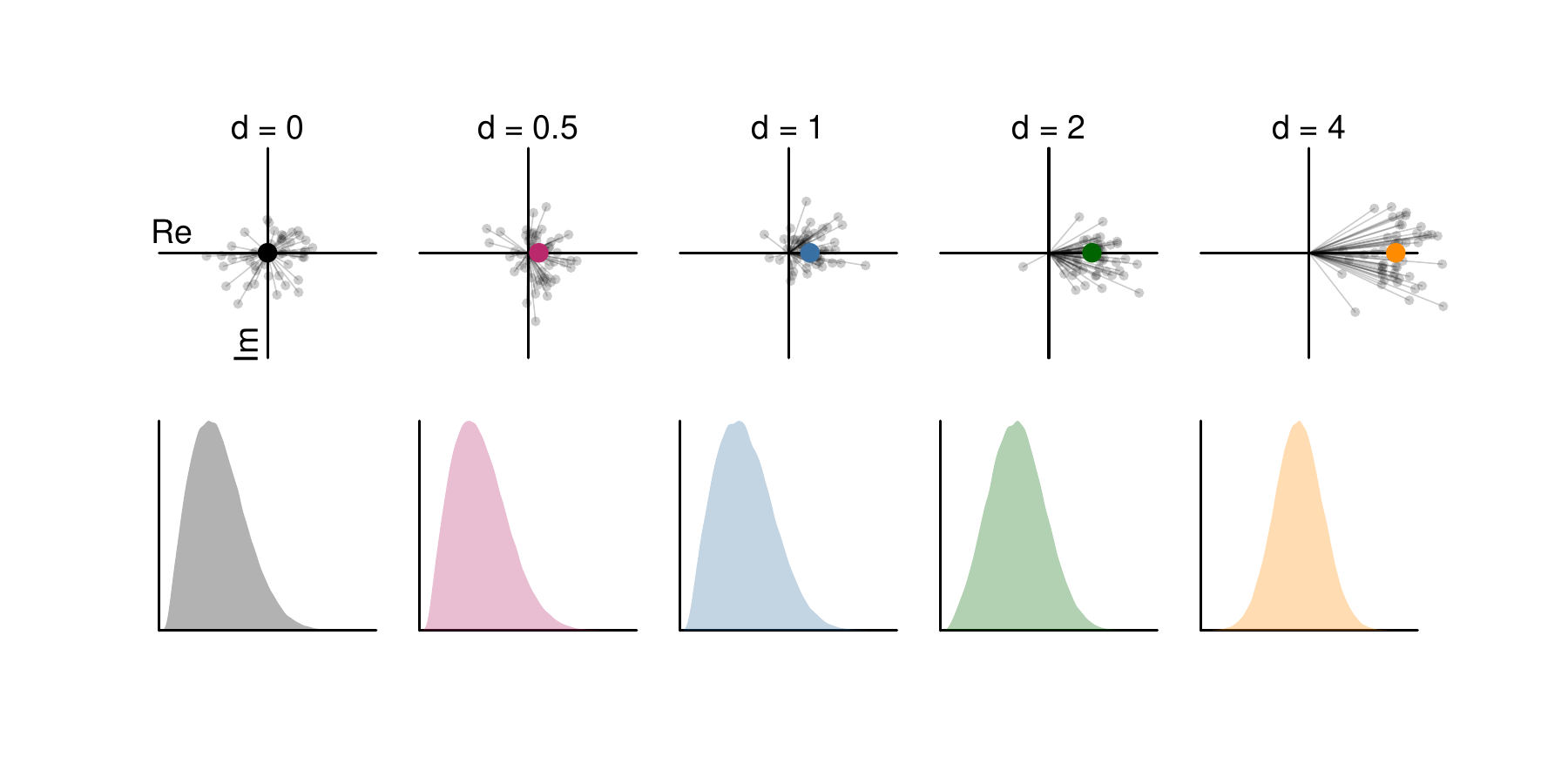} 

}

\caption{Demonstration of skew in absolute Fourier amplitudes for signals of different strengths. Signal strength is quantified as Cohen's \emph{d}, defined as the ratio of the sample mean to the sample standard deviation. The upper row shows samples of 50 grey points, and the population mean (coloured points). The lower row shows kernel density functions generated from 100,000 amplitude values. Note that the mean phase of the signal is irrelevant for these simulations, and is shown in the positive x-direction for consistency.}\label{fig:amphists}
\end{figure}

The upper row of Figure \ref{fig:amphists} shows scatterplots of simulated Fourier components, expressed using real (\emph{x}) and imaginary (\emph{y}) components. The amplitudes are the lengths of the lines joining each grey point to the origin. The lower row in Figure \ref{fig:amphists} shows distributions of amplitudes for the same set of signal strengths. These distributions only approach normality when the signal strength is more than twice the standard deviation (Cohen's \emph{d} \textgreater{} 2; Cohen's \emph{d} is the difference of sample means scaled by the standard deviation, see Cohen \citeyearpar{Cohen1988}). One consequence of this is that T-tests will potentially have an inflated Type I error (false positive) rate for many signals encountered empirically, especially if used to make pointwise comparisons to an amplitude of 0.

Typical solutions for dealing with skew, such as log-transforming the data, are unlikely to be equally applicable to all conditions. For example, if one wishes to compare a baseline where no stimulus was presented with a condition involving a strong signal, the former will be skewed and the latter normal. Applying a transform to both conditions is therefore problematic. Non-parametric statistics are a potential option, but these have generally lower statistical power than their parametric equivalents. Instead, the multivariate statistics discussed in the introduction avoid these issues and have greater statistical power, as will be demonstrated in the following section.

\hypertarget{comparison-of-statistical-power-between-univariate-and-multivariate-tests}{%
\section{Comparison of statistical power between univariate and multivariate tests}\label{comparison-of-statistical-power-between-univariate-and-multivariate-tests}}

Using the \(T^2\) or \(T^2_{circ}\) statistic allows the phase information to be retained, and therefore provides greater power than running univariate T-tests on amplitudes, as well as avoiding problems caused by using absolute amplitude values. \cite{Victor1991} report simulations showing situations where \(T^2_{circ}\) has greater power than \(T^2\), which is expected on theoretical grounds because the additional degrees of freedom in the \(T^2_{circ}\) test make it more efficient. Their simulations involved generating random data sets of different sample sizes, and different signal strengths (sample means), and comparing the number of such tests where each statistic was significant. The simulations showed the largest advantage for \(T^2_{circ}\) for effect sizes around \emph{d} = 1 (where the sample mean is equal to the sample standard deviation). The advantage appeared to be stronger for smaller sample sizes.

\begin{figure}

{\centering \includegraphics[width=0.8\textwidth]{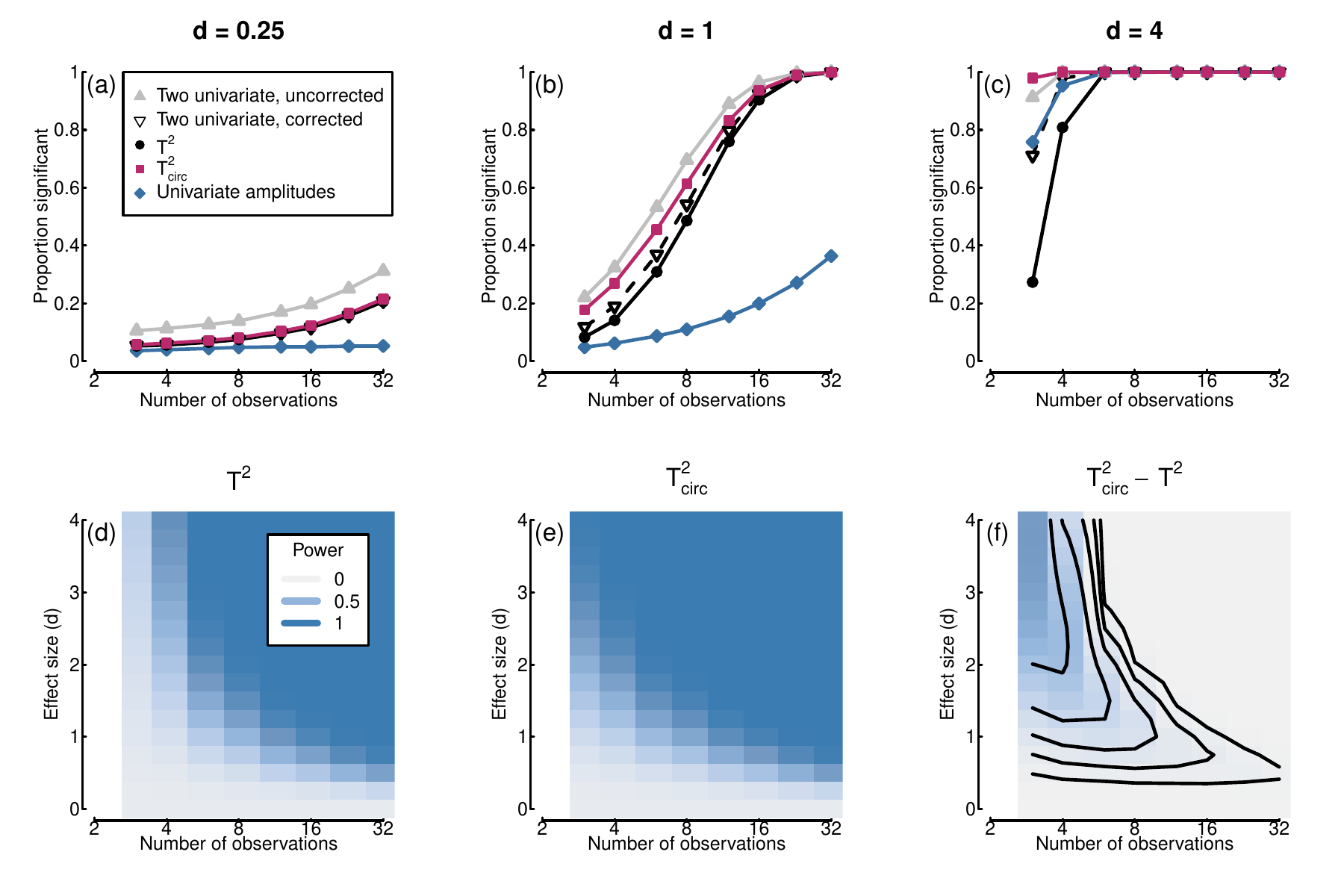} 

}

\caption{Simulations estimating the proportion of significant tests for simulated data with different sample sizes and effect sizes (100,000 simulated data sets per condition). Panels a-c replicate conditions reported by \cite{Victor1991}, and extend them to include univariate tests for comparison. Panels (d) and (e) show a wider range of conditions for each multivariate statistic. Panel (f) shows the difference between the two statistics, with contour lines indicating power differences of 0.02, 0.05, 0.1, 0.2 and 0.4.}\label{fig:powerfig}
\end{figure}

Here these simulations are replicated and extended (see Figure \ref{fig:powerfig}). The real and imaginary Fourier components of a periodic signal could in principle be analysed using two separate univariate t-tests. To detect the presence of a signal, each t-test would compare to a value of zero (i.e.~the origin in Figure \ref{fig:fourierexplain}d), to test against the null hypothesis that there is no signal (\(H_{0x}: x = 0; H_{0y}: y = 0\)). Without correction for familywise error, this approach has high statistical power for detecting when at least one test is significant (see grey triangles in Figure \ref{fig:powerfig}a-c). However the use of two separate tests inflates the familywise error rate such that 10\% of tests are significant (assuming \(\alpha = 0.05\)), even in the absence of a signal. Using Bonferroni correction to ameliorate this problem substantially reduces the power (white triangles in Figure \ref{fig:powerfig}a-c).

Using multivariate tests allows high power to be maintained without inflating the Type I error rate. Here, the multivariate mean (or equivalently, the difference in means between two conditions) is compared to zero, again testing against the null hypothesis that there is no signal (\(H_0: x = y = 0\)). The \(T^2_{circ}\) statistic (red squares in Figure \ref{fig:powerfig}a-c) has greater power than either the \(T^2\) statistic (black circles) or the corrected univariate tests (white triangles). The power advantage over \(T^2\) occurs particularly for large effect sizes and small sample sizes (see Figure \ref{fig:powerfig}f). However, for effect sizes around 0.5 \textless{} \emph{d} \textless{} 1, \(T^2_{circ}\) is more sensitive even with around 16 observations. This advantage is lost for large sample sizes (N \textgreater{} 32) and large effect sizes (when \emph{d} \textgreater{} 2 and N \textgreater{} 8). These simulations suggest a straightforward heuristic - there is no advantage to using the \(T^2_{circ}\) statistic for large sample sizes (when N \textgreater{} 32), so its use can be restricted to small-sample studies.

As mentioned above, an alternative approach that is widely used in the literature is to discard the phase information and calculate the amplitude values instead (lengths of the lines in the top row of Figure \ref{fig:amphists}). Because the amplitudes can never be negative, the presence of a signal must be detected by comparison to a baseline condition in which no stimulus was presented, typically using a paired t-test. Here, the null hypothesis is that the stimulus-present and stimulus-absent conditions produce equal amplitudes at the stimulus frequency (\(H_0: A_{present} = A_{absent}\)). This approach has markedly lower power than the multivariate tests (see blue diamonds in Figure \ref{fig:powerfig}a-c). In principle, one could also conduct circular statistical tests on the phase component of the signal \citep{Berens2009}. However in the limit, testing both the amplitude and phase separately can be no more sensitive than testing the real and imaginary components (because these are just the polar and Cartesian representations of the same information), and suffers the same issues with familywise error from multiple tests. Multivariate approaches avoid these problems and provide high statistical power.

\hypertarget{limitations-of-t2_circ-when-assumptions-are-violated}{%
\section{\texorpdfstring{Limitations of \(T^2_{circ}\) when assumptions are violated}{Limitations of T\^{}2\_\{circ\} when assumptions are violated}}\label{limitations-of-t2_circ-when-assumptions-are-violated}}

\begin{figure}
\centering
\includegraphics[width=0.95\textwidth]{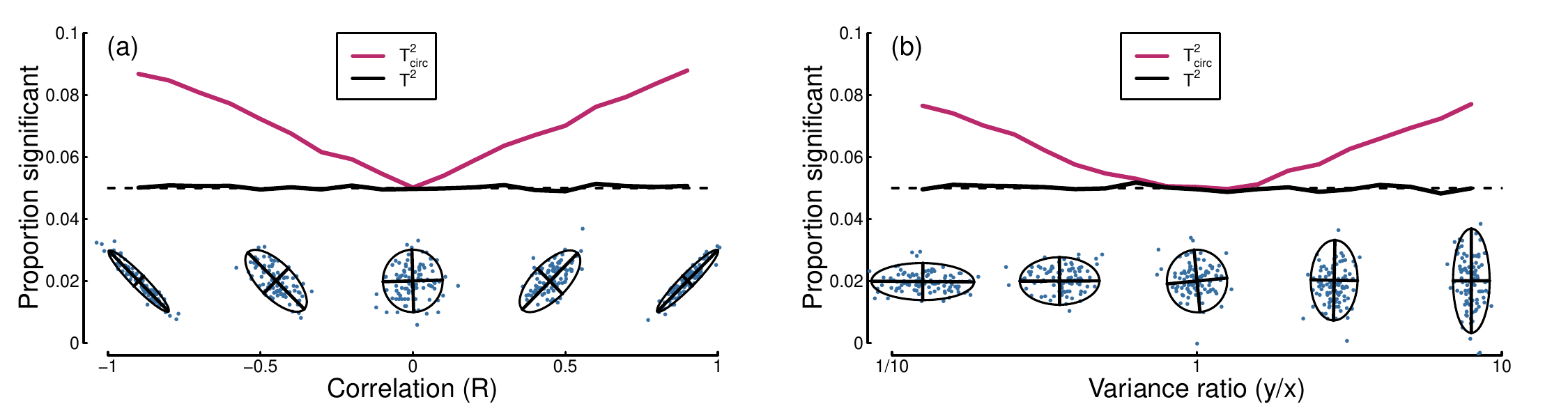}
\caption{\label{fig:falsealarms}Simulations showing the Type I error rate for both tests as a function of the correlation between two variables (a) and the ratio of variances (b). Estimates are for 100,000 simulated data sets per condition, with N = 10 observations. The icons at the foot of each panel show example scatterplots with bounding ellipses and eigenvectors.}
\end{figure}

Although \(T^2_{circ}\) can be more sensitive than \(T^2\), this greater sensitivity relies on satisfying the \(T^2_{circ}\) test's more stringent assumptions. The two dependent variables must be independent (i.e.~uncorrelated), and of equal variance. These restrictions may hold for some data sets, but it is instructive to ask what happens when they do not. Figure \ref{fig:falsealarms} shows the results of simulations with randomly generated bivariate data in which no signal is present. When the data are uncorrelated and have equal variance (mid-points of the functions in each panel), both tests have the nominal Type I error (false positive) rate of \(\alpha = 0.05\) (horizontal dashed lines). However, as the data become increasingly correlated (Figure \ref{fig:falsealarms}a), or the sample variances of the two dependent variables grow more disparate (Figure \ref{fig:falsealarms}b), the Type I error rate of the \(T^2_{circ}\) statistic (shown in red) increases by almost a factor of 2. In contrast, the \(T^2\) statistic, which explicitly takes account of the covariance matrix (see equation \eqref{eq:t2eq}) shows no increase (black curves).

One possible remedy to control the Type I error rate would be to adjust either the \(\alpha\) level or the degrees of freedom (as is done in repeated measures ANOVA when sphericity assumptions are violated). However, this will reduce the statistical power of the \(T^2_{circ}\) test, and its advantage over \(T^2\) is relatively marginal in most situations to begin with (see Figure \ref{fig:powerfig}). What is required is a method to objectively assess whether the assumptions of \(T^2_{circ}\) hold; this is developed in the following section.

\hypertarget{a-novel-method-to-test-the-assumptions-of-t2_circ}{%
\section{\texorpdfstring{A novel method to test the assumptions of \(T^2_{circ}\)}{A novel method to test the assumptions of T\^{}2\_\{circ\}}}\label{a-novel-method-to-test-the-assumptions-of-t2_circ}}

Despite the severe consequences of violating the assumptions of the \(T^2_{circ}\) statistic (see Figure \ref{fig:falsealarms}), there is currently no accepted test of those assumptions that could be applied to an empirical data set. \cite{Victor1991} suggest that their test should be applicable to multiple repetitions of a stimulus condition collected from a single participant, whereas data pooled across multiple participants may be less likely to exhibit independence of the real and imaginary components \citep[see also][]{Pei2017}. However it would be useful to develop a method that can tell us whether the assumptions hold for a given data set.

\begin{figure}

{\centering \includegraphics[width=0.8\textwidth]{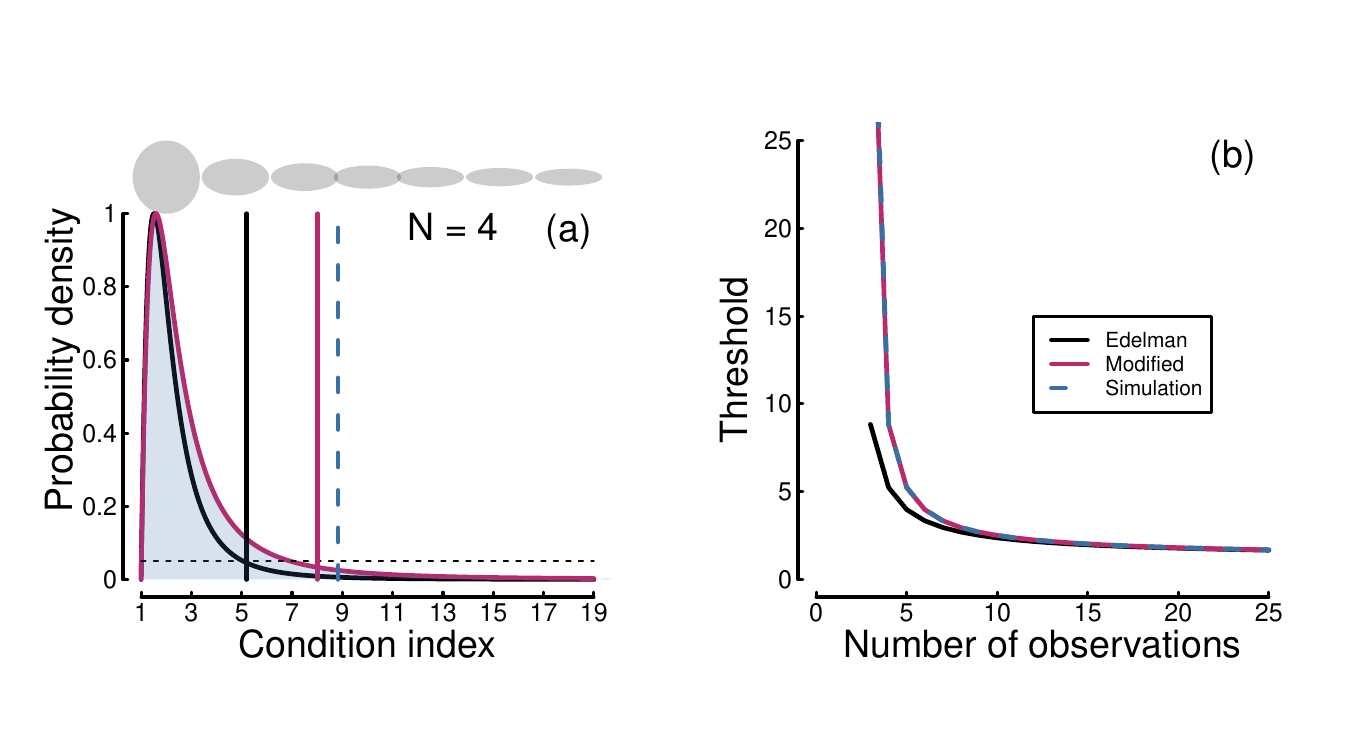} 

}

\caption{Logic of the condition index test. Panel (a) shows the distribution of condition indices derived from Equations 3 (black curve) and 4 (red curve), and by stochastic simulation (blue shading), for a sample size of N = 4 observations. The vertical lines show the 95 percent thresholds on the distributions (where 95 percent of values lie to the left of the line). Ellipse icons above panel (a) illustrate different condition indices between 1 and 19 (note that rotation does not affect the condition index). Panel (b) shows how 95 percent thresholds change as a function of the number of observations.}\label{fig:distcomparison}
\end{figure}

One convenient way to test the assumptions of \(T^2_{circ}\) is to assess the \emph{condition index} of a data set, which describes the ratio of eigenvalues for a cloud of points. The eigenvectors are the major and minor axes of the bounding ellipse (the straight lines in the example icons at the foot of Figure \ref{fig:falsealarms}a,b). Conventionally, the condition index is calculated as the square root of the longest/shortest eigenvector length. For uncorrelated random numbers the expected distribution of condition indices is positively skewed, with a minimum of 1 \citep{Edelman1988}. This is because two independent samples of numbers from the same underlying distribution will generally by chance have unequal eigenvectors, and the definition of the condition index (\(\sqrt{longest/shortest}\)) prevents its values dropping below 1 (values \textless{} 1 would imply that the shortest eigenvector is longer than the longest one). For bivariate data, \cite{Edelman1988} provides an equation (his Eq. 14) for the probability density function of condition indices (\emph{x}) as a function of sample size (\emph{N}):
\begin{equation}
\label{eq:edelman1}
pdf = (N-1)2^{N-1}\frac{x^2 - 1}{(x^2 + 1)^N}x^{(N-2)}.
\end{equation}
Attempts to validate this by simulation (100,000 random data sets) found that for small sample sizes (\emph{N} \textless{} 10), there is a mismatch between the equation's predictions and the simulations (the black curve does not match the blue shading in Figure \ref{fig:distcomparison}a). Instead, a closer approximation to the simulations is given by:
\begin{equation}
\label{eq:edelman2}
pdf = (N-2)2^{N-2}\frac{x^2 - 1}{(x^2 + 1)^{(N-1)}}x^{(N-3)}.
\end{equation}
The red curve in Figure \ref{fig:distcomparison}a shows the predictions of this modified equation, which are much closer to the simulation results (blue shading). The vertical lines show the critical (95\%) threshold for the analytic and simulated results. A ratio lying beyond this threshold can be considered to violate the assumption of either independence or equal variance, because it has a condition index larger than expected by chance (assuming \(\alpha\) = 0.05). Ratios below the threshold imply that the eigenvalues can be considered equal (in a statistical sense). Figure \ref{fig:distcomparison}b shows how these critical thresholds change as a function of the number of observations (N), and it appears that the modified expression (red) most closely approximates the simulation results (blue).

The eigenvalue ratio can be used as a test of the assumptions of \(T^2_{circ}\). If a condition index is observed that is above the critical threshold for the number of observations, then the data set can be said to significantly violate the assumption of equal eigenvalues. Because the modified equation permits estimation of an inverse density function, this can be used to calculate a \emph{p}-value for the test. If the test is non-significant, one can proceed with \(T^2_{circ}\); if it is significant, \(T^2\) should be used instead. A function implementing this test is included in the \emph{FourierStats} package (the \emph{CI.test} function in \emph{R}, and \emph{CI\_test} function in \emph{Matlab}).

\hypertarget{identifying-and-removing-outliers-using-the-mahalanobis-distance}{%
\section{Identifying and removing outliers using the Mahalanobis distance}\label{identifying-and-removing-outliers-using-the-mahalanobis-distance}}

If a data set produces a significant result using the condition index test, this could be due to the presence of one or more outliers. The Mahalanobis distance \citep{Mahalanobis1936} is a useful metric for identifying such multivariate outliers so that they can be excluded. Note that outlier exclusion is contentious, but it is beyond the scope of this paper to contribute to this debate beyond noting that preregistration of any outlier exclusion protocols is adviseable. The Mahalanobis distance calculates the Euclidean distance between each data point and the sample mean, and scales it by the variance in the direction of the vector that joins the two points. This means that any correlations in the data set are taken into account when calculating the distance metric, \emph{D}.

\begin{figure}

{\centering \includegraphics[width=0.6\textwidth]{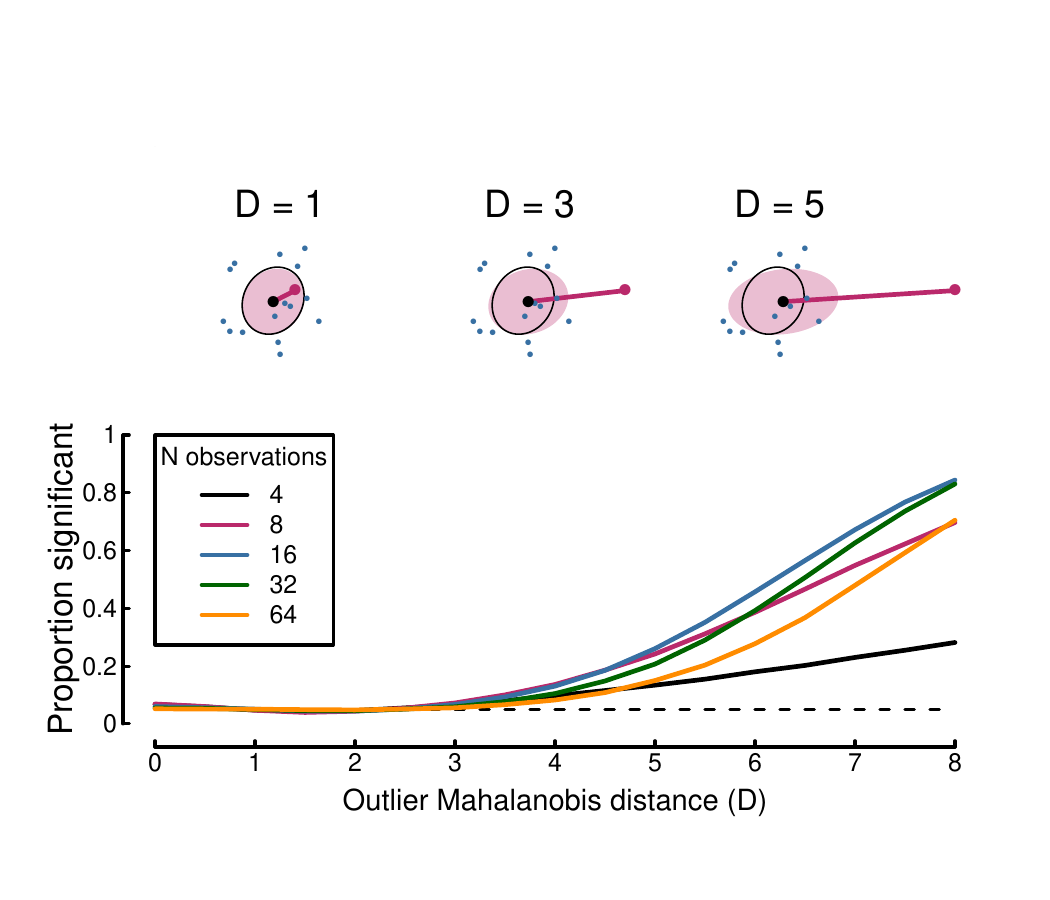} 

}

\caption{Simulations illustrating the Mahalanobis distance metric, and showing how a single outlier affects the condition index. The upper row shows three example data sets, each with a single outlier shown in red. The outliers have approximate Mahalanobis distances of 1, 3 and 5. The ellipses are calculated with the outlier included (red) and excluded (black), illustrating how the outlier distorts the aspect ratio of the ellipse. The main plot shows how the proportion of significant condition index tests depends on the outlier distance and the sample size.}\label{fig:outlierplot}
\end{figure}

The effectiveness of this approach to outlier exclusion can be assessed by simulation using the condition index test. Figure \ref{fig:outlierplot} shows the proportion of significant condition index tests as a function of the Mahalanobis distance of a single outlier, for a range of sample sizes (curves). In all cases, the functions depart from the Type I error rate (\(\alpha\) = 0.05; horizontal dashed line in Figure \ref{fig:outlierplot}) when the outlier's Mahalanobis distance exceeds a value around 3. This seems a reasonable heuristic for outlier exclusion, and is the multivariate equivalent of excluding data points more than 3 standard deviations from the sample mean (note that many implementations of the Mahalanobis distance statistic, such as the core \emph{mahalanobis} function in \emph{R}, or \emph{mahal} function in \emph{Matlab}, return \(D^2\), which can be converted to \emph{D} by taking the square root). Following this heuristic should reduce the likelihood that outliers will invalidate the assumptions of the \(T^2_{circ}\) test.

A variant of the Mahalanobis distance (the pairwise Mahalanobis distance) can also be used to compute a multivariate measure of effect size, equivalent to Cohen's \emph{d} statistic \citep[see e.g.][]{Giudice2009}. This is a valuable statistic to include when reporting the results of multivariate tests, and a function (\emph{pairwisemahal}) to calculate it is available as part of the \emph{FourierStats} package.

\hypertarget{controlling-for-multiple-comparisons-across-location-and-time}{%
\section{Controlling for multiple comparisons across location and time}\label{controlling-for-multiple-comparisons-across-location-and-time}}

In some studies, it is important to compare responses over space and/or time. However with large numbers of sensors, voxels or temporal epochs, the familywise error quickly becomes problematic, inflating the Type I error (false positive) rate. Solutions such as Bonferroni correction, which adjust the \(\alpha\) level based on the number of comparisons, are overly conservative and can obscure real effects by dramatically reducing power. An alternative approach is to use cluster correction methods to control the Type I error rate, using mass univariate tests \citep[e.g.][]{Maris2007}. These typically involve summing test statistics such as T-values across adjacent significant locations and/or moments in time. The summed test statistic is compared to a null distribution generated from the same data by randomly permuting condition labels (or the sign of the data for one-sample tests). Such methods control the Type I error rate without substantially reducing statistical power. The same approach can be applied to the \(T^2\) and \(T^2_{circ}\) statistics. This allows a principled method for identifying clusters of significant sensors, timepoints or frequencies responding to periodic stimuli. The \emph{FourierStats} package includes an implementation of this method with options for multivariate statistics (the \emph{clustercorrect} function).

\hypertarget{generalising-to-more-than-two-conditions}{%
\section{Generalising to more than two conditions}\label{generalising-to-more-than-two-conditions}}

Many studies involve more than two experimental conditions that need to be compared. Again, issues with familywise error will quickly become problematic if multiple pairwise \(T^2\) or \(T^2_{circ}\) statistics are calculated. One possibility is to conduct a MANOVA, which takes covariances between dependent variables into account in much the same way as Hotelling's \(T^2\), but permits independent variables with more than two levels, as well as factorial designs. However, if the assumptions of \(T^2_{circ}\) hold for a data set, it should alternatively be possible to extend the logic of the \(T^2_{circ}\) test \citep{Victor1991} to the more general multiple group case, and obtain a sensitivity benefit similar to that shown in Figure \ref{fig:powerfig}.

The F-statistic for a one-way independent ANOVA is calculated by taking the ratio between the variance explained by the modelled group means, and the residual unexplained variance:
\begin{equation}
\label{eq:Fratio}
F = \frac{MS_M}{MS_R},
\end{equation}
where \(MS_M\) is the mean squares of the linear model, and \(MS_R\) is the mean squares of the residuals. For bivariate data, the linear model is defined as the change in group (condition) means relative to the grand mean, calculated using the vector distances between points:
\begin{equation}
\label{eq:MSM}
MS_M = \frac{\Sigma N_k (\bar{x}_k - \bar{x}_{grand})^2}{df_M},
\end{equation}
where \(N_k\) is the number of observations in group \emph{k}, \(\bar{x}_k\) is the bivariate sample mean of group \emph{k}, \(\bar{x}_{grand}\) is the bivariate mean of the entire sample (including all groups), \(\Sigma\) denotes summation across groups, and \(df_M\) is the degrees of freedom for the model. Note that by using the vector distances between points, this equation reduces multivariate data to a single scalar value, in a similar way to equations \eqref{eq:t2eq} and \eqref{eq:t2c}. The model degrees of freedom for \emph{k} groups is \(df_M = 2(k-1)\), with the factor of two scaling relative to standard univariate ANOVA reflecting the additional degrees of freedom afforded by having two dependent variables.

The residuals are the vector distances between each data point and its corresponding group mean. The denominator component for the F-ratio equation is therefore defined as:
\begin{equation}
\label{eq:MSR}
MS_R = \frac{\Sigma (x_{i,k} - \bar{x}_k)^2}{df_R},
\end{equation}
where \(x_{i,k}\) is the \(i\)th data point from group \emph{k}, and other terms are as described previously. The degrees of freedom for a balanced design are calculated as \(df_R = 2((Nk)-k)\), where \emph{N} is the number of observations per group, and \emph{k} is the number of groups. A \emph{p}-value can then be determined by comparing the F-ratio calculated from equation \eqref{eq:Fratio} to an F-distribution with \(df_M\) and \(df_R\) degrees of freedom.

\begin{figure}

{\centering \includegraphics[width=0.8\textwidth]{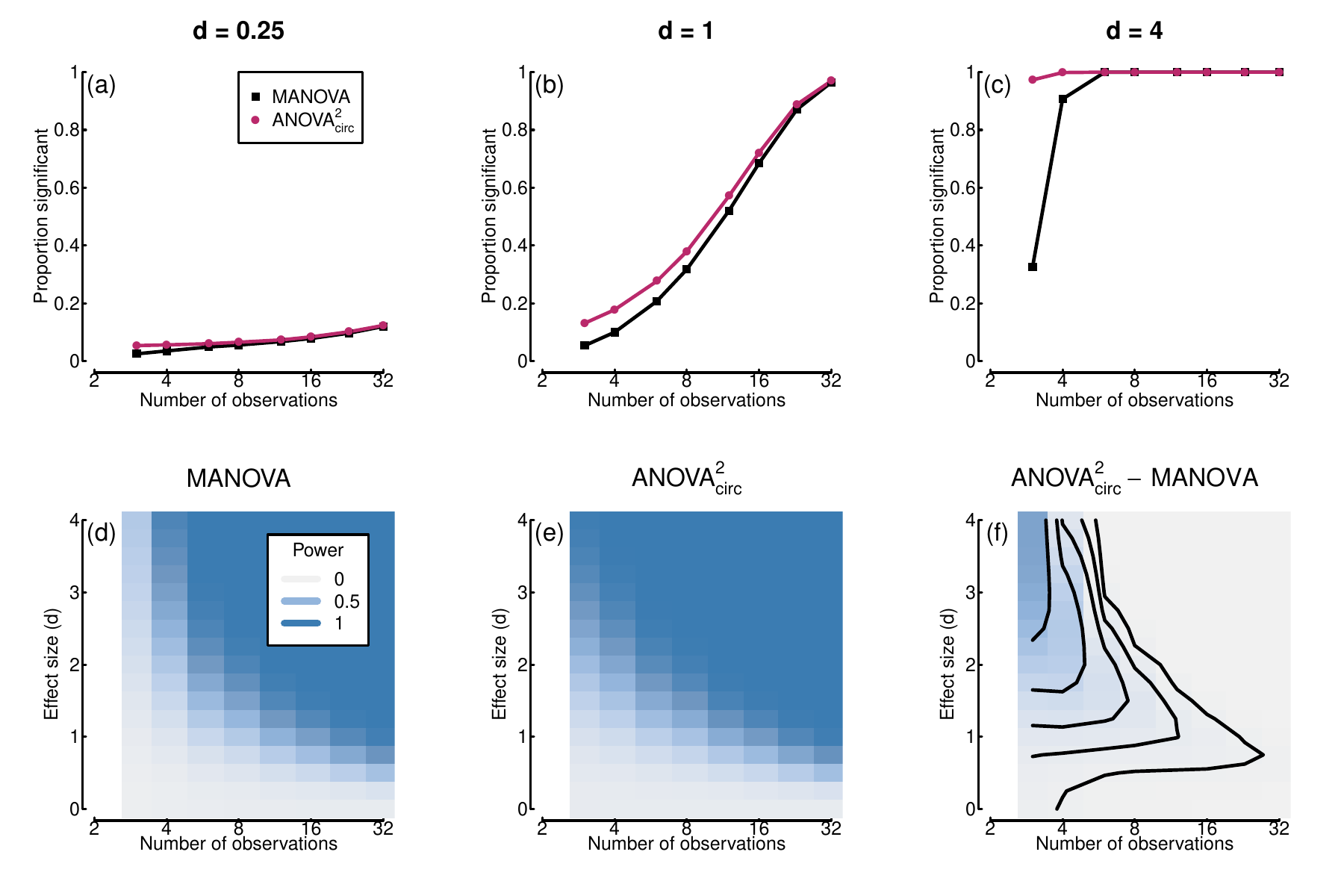} 

}

\caption{Simulations comparing the sensitivity of MANOVA and \({ANOVA}^2_{circ}\). The format mirrors that of Figure 3. In these simulations, there were three conditions, with the signal being added to one condition only.}\label{fig:powerfig2}
\end{figure}

A suitable name for such a test might be \({ANOVA}^2_{circ}\), as this reflects the similarity to ANOVA, and the extension of the logic of \(T^2_{circ}\) (an alternative name might be \(MANOVA_{circ}\), however this feels less appropriate given that many of the key features of MANOVA are absent). Figure \ref{fig:powerfig2} shows simulations analogous to those in Figure \ref{fig:powerfig} for a one-way between-subjects design with three levels. MANOVA is directly compared to the \({ANOVA}^2_{circ}\) statistic across a range of effect sizes and sample sizes. Just as for the one-sample statistics, the advantages of \({ANOVA}^2_{circ}\) are particularly apparent for small sample sizes, and larger effect sizes (Figure \ref{fig:powerfig2}f).

Following a significant \({ANOVA}^2_{circ}\) test, one could calculate \(T^2_{circ}\) statistics to make post-hoc pairwise comparisons between conditions, providing that appropriate multiple comparison correction is applied (e.g.~Bonferroni correction). Univariate ANOVAs on the real and imaginary components are unlikely to be informative, as the relative magnitudes of the two measures depend on stimulus phase (which is arbitrary; see Figure \ref{fig:fourierexplain}d). A repeated measures version of \({ANOVA}^2_{circ}\) can also be implemented following the same logic. The project repository contains a function (\emph{anovacirc.test} in \emph{R}, \emph{anovacirc\_test} in \emph{Matlab}) to run both of these tests for one-way designs. In principle factorial versions might also be derived.

\hypertarget{deciding-which-test-to-run}{%
\section{Deciding which test to run}\label{deciding-which-test-to-run}}

\begin{figure}

{\centering \includegraphics[width=0.8\textwidth]{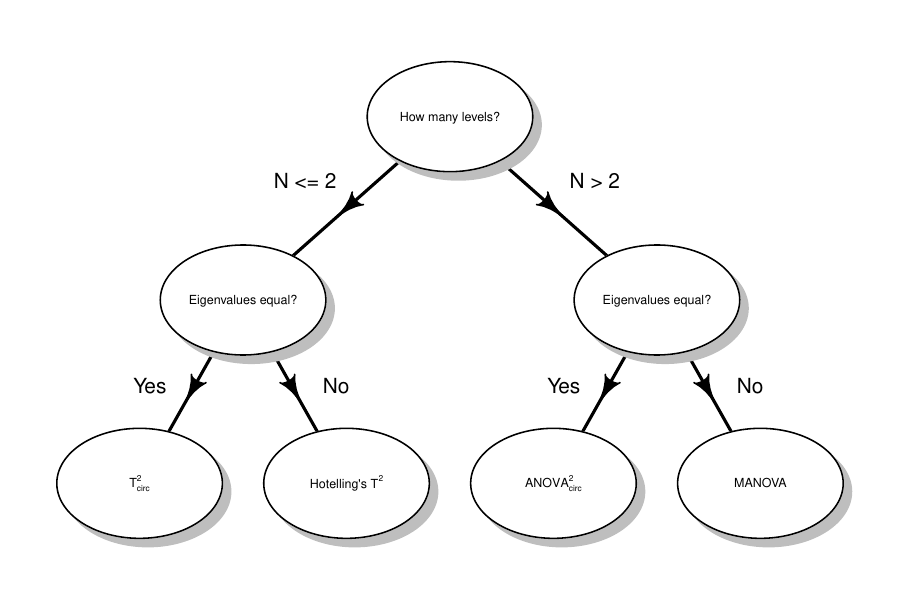} 

}

\caption{Flowchart illustrating how one might decide which test to conduct for a given data set, based on the study design and the outcome of the condition index test.}\label{fig:flowchart}
\end{figure}

The flowchart in Figure \ref{fig:flowchart} illustrates a proposed decision structure for the analysis of periodic data, once any outliers have been removed. Initially, data for each condition should be tested against the expected distribution of eigenvalue ratios using the condition index test. Comparisons with one or two conditions should be tested with the \(T^2_{circ}\) statistic if the condition index test is non-significant, and the \(T^2\) statistic otherwise. Comparisons with more than two conditions should be tested with the \(ANOVA^2_{circ}\) statistic if the condition index test is non-significant, or a MANOVA otherwise. Many MANOVA implementations cannot deal correctly with random factors (repeated measures), particularly in complex factorial designs. However the \emph{multRM} function in the \href{https://CRAN.R-project.org/package=MANOVA.RM}{\emph{MANOVA.RM}} package \citep{Friedrich2019} is able to handle such designs appropriately using a bootstrapping approach.

\hypertarget{applying-multivariate-methods-to-empirical-data-sets}{%
\section{Applying multivariate methods to empirical data sets}\label{applying-multivariate-methods-to-empirical-data-sets}}

Having developed some novel tools for the analysis of periodic data, in the following sections their use is demonstrated for two different publicly available empirical data sets. The first study recorded responses to auditory and optogenetic stimulation in mice. The second study measured visual responses to flickering grating patterns in humans. These examples also provide a demonstration of how the results of the tests described above might be appropriately reported.

\hypertarget{mouse-auditory-and-optogenetic-steady-state-data}{%
\subsection{Mouse auditory and optogenetic steady-state data}\label{mouse-auditory-and-optogenetic-steady-state-data}}

Hwang et al. \citeyearpar{Hwang2019, Hwang2020} measured steady-state responses using implanted scalp electrodes in 6 mice. The mice had previously been given a targeted virus that made parvalbumin neurons in their basal forebrain responsive to specific wavelengths of light, delivered through an optical fiber (a technique called optogenetics). Steady-state evoked potentials were recorded from 36 electrodes for 1 second epochs of 40 Hz auditory stimulation, and various schedules of optogenetic stimulation (including at 40 Hz). The data set is described more fully by \cite{Hwang2020}, and was downloaded from: \url{https://doi.gin.g-node.org/10.12751/g-node.e5tyek/}. A processed data file is included with permission from the authors in the \emph{FourierStats} package.

\begin{figure}

{\centering \includegraphics[width=0.7\textwidth]{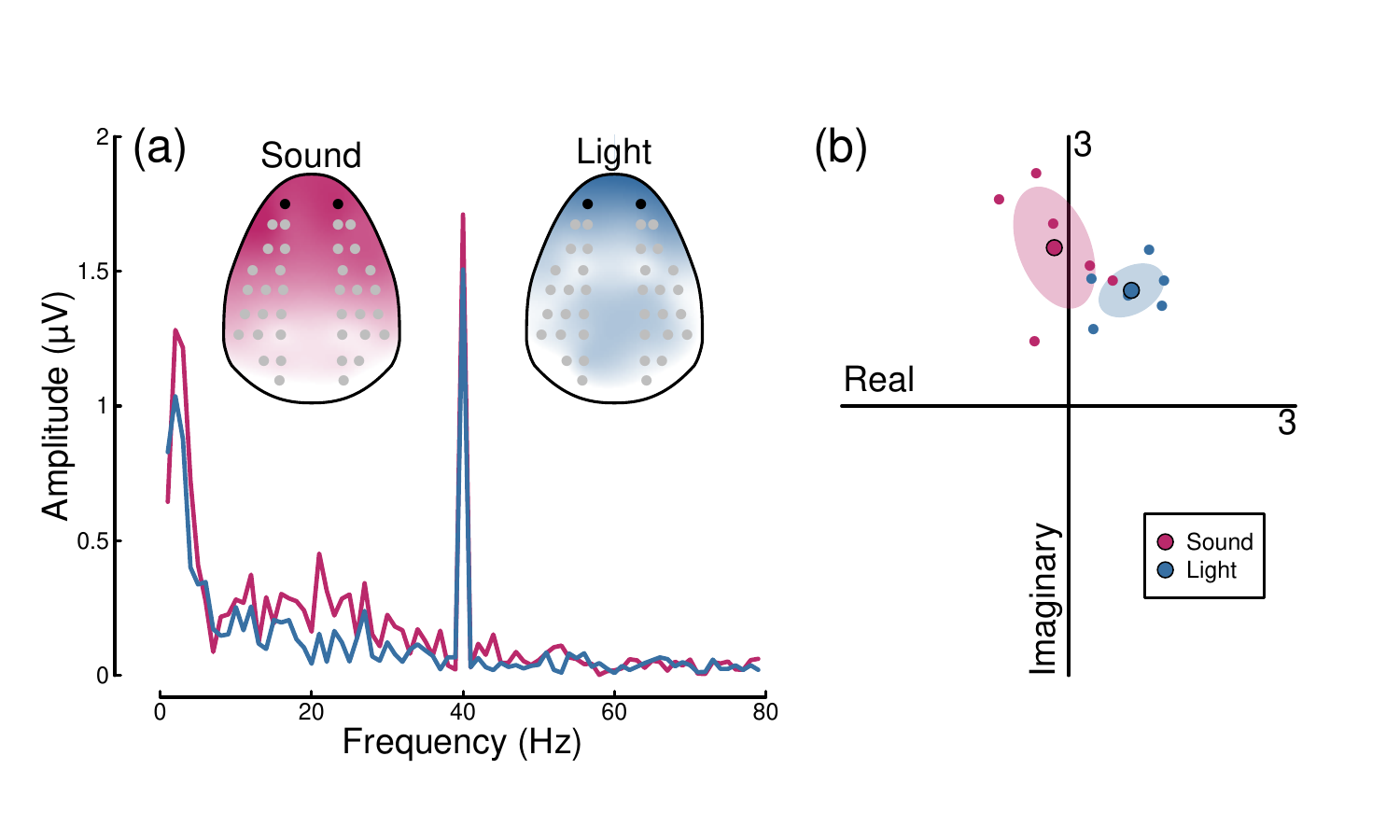} 

}

\caption{Summary of mouse steady-state responses to 40 Hz stimulation. Panel (a) shows the Fourier amplitude spectrum with inset scalp plots for sound (red) and light (blue) stimulation, averaged across repetitions and individuals. Grey and black points in the insets indicate electrode locations. Panel (b) shows complex (\emph{x} = real, \emph{y} = imaginary) Fourier components for 6 individual mice (small points) and their average (large points) for both conditions.}\label{fig:mousedata}
\end{figure}

Figure \ref{fig:mousedata}a shows Fourier amplitude spectra at two frontal electrodes (marked black in the insets) for auditory stimulation (red) and optogenetic stimulation (blue), each at 40 Hz. There is a clear frequency-locked signal with approximately equal amplitude for each stimulation modality. Indeed, a paired univariate T-test on the amplitudes reveals no significant difference (\emph{t} = 0.84, \emph{df} = 5, \emph{p} = 0.44). However, inspection of the complex Fourier components for each condition suggests evidence of a phase difference between the two modalities (see Figure \ref{fig:mousedata}b). The condition index test was non-significant for both conditions (sound: CI = 1.59, \emph{p} = 0.66; light: CI = 1.69, \emph{p} = 0.59), so a paired-samples \(T^2_{circ}\) test was conducted. This revealed a significant difference between conditions (\(T^2_{circ}\) = 1.39, \(F_{(2,10)}\) = 8.32, \emph{p} = 0.007) with an effect size of \emph{D} = 2.14. This demonstrates that sound and (optogenetic) light entrain neural responses with different latencies (lags). The original study by \cite{Hwang2019} went on to explore interactions between these two signals.

\hypertarget{human-visual-steady-state-data}{%
\subsection{Human visual steady-state data}\label{human-visual-steady-state-data}}

\begin{figure}

{\centering \includegraphics[width=0.99\textwidth]{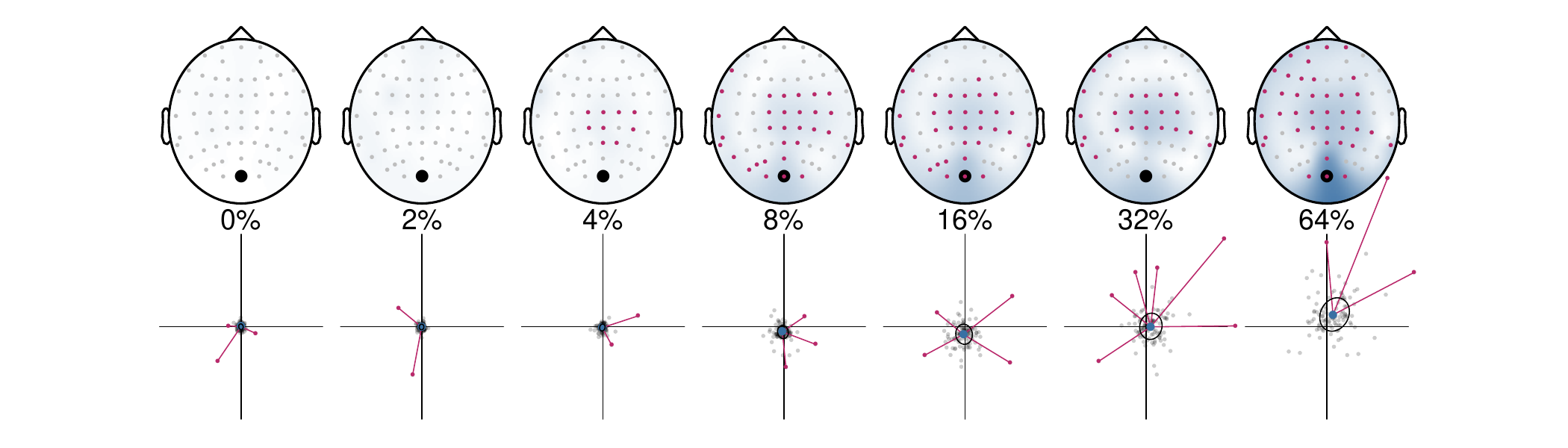} 

}

\caption{Summary of human SSVEP data. Upper row shows scalp distributions of Fourier amplitudes at 7Hz for stimuli of increasing contrasts (blue shading indicates higher amplitudes). Electrodes marked in red indicate cluster-corrected significance.  Lower row shows scatterplots of complex (\emph{x} = real, \emph{y} = imaginary) Fourier components for 100 participants per condition, from electrode \emph{Oz} (marked  by the black points in the scalp plots). Red points are outliers with Mahalanobis distances exceeding 3, and blue points mark the bivariate sample means.}\label{fig:humanSSVEP}
\end{figure}

\cite{Vilidaite2018} measured visual responses to flickering grating stimuli in a large sample of 100 adults. Each participant completed a series of 11-second trials, in which stimuli of different contrasts flickered at 7Hz (on-off sinusoidal flicker). Responses were strongest at occipital electrodes over visual cortex (see upper row of Figure \ref{fig:humanSSVEP}), were well-isolated in the Fourier domain, and increased with stimulus contrast. Significant activity was evident at 4\% contrast and above following cluster correction (with very stringent \(\alpha\) levels given the high power of this data set), as indicated by the red electrodes in the upper row of Figure \ref{fig:humanSSVEP}. For the main analysis, responses were taken from electrode \emph{Oz} at the occipital pole (black points in the upper row of Figure \ref{fig:humanSSVEP}), and averaged across repetition for each participant. Each condition included some outlier points with Mahalanobis distances exceeding 3, marked red in the lower row of Figure \ref{fig:humanSSVEP}. Any participant that contributed at least one outlier was excluded, leaving a total of 89 participants for the main analysis.

With the outlier points removed, all seven conditions resulted in non-significant condition index tests (largest CI = 1.20, all \emph{p} \textgreater{} 0.23). A repeated measures \(ANOVA^2_{circ}\) test was conducted, revealing a significant effect of stimulus contrast (\(F_{(12,1056)}\) = 38.9, \emph{p} \textless{} 0.001). Pairwise \(T^2_{circ}\) statistics comparing the baseline (0\% contrast) condition to each subsequent condition (Bonferroni corrected for 6 tests to \(\alpha = 0.008\)) revealed significant differences at 8\% contrast (\(T^2_{circ}\) = 0.32, \(F_{(2,176)}\) = 28.43, \emph{D} = 1.1, \emph{p} \textless{} 0.001), 16\% contrast (\(T^2_{circ}\) = 0.28, \(F_{(2,176)}\) = 25.25, \emph{D} = 1.04, \emph{p} \textless{} 0.001), 32\% contrast (\(T^2_{circ}\) = 0.10, \(F_{(2,176)}\) = 8.55, \emph{D} = 0.63, \emph{p} \textless{} 0.001) and 64\% contrast (\(T^2_{circ}\) = 0.40, \(F_{(2,176)}\) = 35.35, \emph{D} = 1.17, \emph{p} \textless{} 0.001). The study by Vilidaite et al. (2018) compared SSVEP responses between individuals with and without autism, as well as in a \emph{Drosophila} genetic model of developmental disorders. The raw data are available at: \url{http://dx.doi.org/10.17605/OSF.IO/Y4N5K}, and a processed version is included with the \href{https://github.com/bakerdh/FourierStats}{\emph{FourierStats}} package.

\hypertarget{further-considerations}{%
\section{Further considerations}\label{further-considerations}}

It is worth stating explicitly that the statistical tests discussed in this paper are applicable only when the signal phase is expected to be consistent across observations. This is the case for most paradigms in which the nervous system is driven by a periodic stimulus, and sometimes also when phase-locked oscillatory responses are induced by a brief stimulus presentation \citep[e.g.~in time-frequency analysis of event-related potentials, see][]{Delorme2004, Pfurtscheller1979}. However, the methods are less obviously applicable to the analysis of endogenous neural oscillations and brain rhythms \citep{Berger1929, Buzsaki2004}, which will typically have random phase and broader bandwidths in the Fourier domain: other analysis methods have been developed for such signals \citep[e.g.][]{Quinn2021}. When phases are consistent across repetitions, greater statistical power can be obtained by coherently averaging across repetitions to obtain a participant-level average \citep{Baker2021}. This practice is also necessary in order to use the multivariate methods discussed here, as the alternative is to discard the phase information and average amplitudes instead, rendering the data univariate.

The present paper has focussed on the Frequentist statistical tradition. However there are many advantages to the Bayesian approach, in which one can make direct quantitative comparisons of the evidence supporting both the experimental and null hypotheses \citep{Jeffreys1961}. Subject to determining appropriate priors, Bayes factor scores might be calculated for Bayesian versions of all of the statistics considered here, much as has been done previously for univariate T-tests \citep{Rouder2009} and ANOVA \citep{Rouder2017}. However this is a non-trivial undertaking, and is beyond the scope of the current paper.

Another possibility is to use machine learning techniques such as multivariate pattern analysis (MVPA) to analyse periodic data. This involves training a classifier algorithm to distinguish between two (or more) experimental conditions or states, and then assessing classifier accuracy for predicting the group labels of fresh data. If different conditions produce distinct patterns of neural response, then classifier accuracy will be above chance. Such methods have been hugely influential in the fMRI literature \citep{Schwarzkopf2011}, and for analysing event-related potential data collected using EEG or MEG \citep{Grootswagers2017}. However, they have not been widely applied to steady-state data \citep[though see][for one example]{West2015}. In principle, the real and imaginary Fourier components can be treated as separate dependent variables, along with different recording locations and/or frequencies. This approach has the potential to offer sensitive, high-powered statistical tests that circumvent many of the shortcomings associated with traditional statistics.

Even when statistics are conducted using both the real and imaginary Fourier components, it is still typical to visualise the mean amplitudes. Several approaches to calculating appropriate error bars have been proposed. For example, \cite{Pei2017} suggest calculating the nearest and farthest points from the origin on the bounding ellipse, and using these to derive standard errors for the amplitude. This approach is somewhat computationally demanding, though a function is provided to calculate error bars using this method (available through the \emph{amperrors} function). However, as an alternative, bootstrap resampling offers a powerful and general method for calculating confidence intervals on amplitudes. This is achieved by resampling the complex data (with replacement) and calculating a resampled complex mean. The amplitude is then derived for this resampled mean, and the procedure repeated a large number of times (1000 or 10000 repetitions is typical) to build up a population of resampled mean amplitudes. Upper and lower confidence intervals on the amplitude can then be taken at appropriate quantiles of this population (68\% or 95\% are typical, with the 68\% interval corresponding to the standard error).

\hypertarget{general-recommendations-for-analysing-periodic-data}{%
\section{General recommendations for analysing periodic data}\label{general-recommendations-for-analysing-periodic-data}}

The simulations reported here allow several recommendations to be made for how periodic data should be analysed. Multivariate statistics should be used for phase-locked Fourier data instead of univariate statistics such as T-tests and ANOVA. This avoids problems from non-normal distributions of amplitudes violating the test assumptions, and also provides a sensitivity benefit from the inclusion of phase information. Outliers can be removed when they have a Mahalanobis distance exceeding 3, and the pairwise Mahalanobis distance should be reported as a measure of effect size. For sample sizes of N \textless{} 32, the \(T^2_{circ}\) and \(ANOVA^2_{circ}\) statistics can be used if the condition index test is non-significant for all conditions. Alternatively, the \(T^2\) or MANOVA statistics should be used when these conditions are not met. The greater power afforded by these tests should in general lead to more accurate statistical inferences when analysing periodic data. Additionally, more sensitive methods require fewer observations to reach a given level of power, which has important ethical implications for animal research if fewer experimental subjects are needed.

\section*{}
\bibliography{bibliography}

\end{document}